\documentclass[twocolumn,showpacs,amsmath,amssymb,floatfix,prl,aps,superscriptaddress,showkeys]{revtex4}

\usepackage{graphicx}% Include figure files
\usepackage{epsfig}
\usepackage{dcolumn}% Align table columns on decimal point
\usepackage{bm}% bold math
\usepackage{times}

\begin{document}
\title{Optical $\Lambda$ transitions and quantum computing in the $^{15}$N-V$^{-}$ Center in Diamond}

\author{Gabriel Gonz\'alez and Michael N. Leuenberger}\email{mleuenbe@mail.ucf.edu}
\address{NanoScience Technology Center, University of Central
Florida, Orlando, FL 32826, USA}
\address{Department of Physics, University of Central
Florida, Orlando, FL 32816-2385, USA}

\begin{abstract}
We present a thorough derivation of the excited state energy levels of the negatively charged $^{15}$N-V$^{-}$ center in diamond for the strong applied electric field case. We show that in the $^{15}$N-V$^{-}$ center a spin non-conserving two-photon $\Lambda$ transition exists that is mediated by the hyperfine interaction, which provides the possibility to write quantum information. Using second order perturbation theory we  obtain a $\Lambda$ transition rate of the order of 10 MHz at room temperature, which allows for approximately $10^4$ quantum logic operations within the spin coherence time $\tau_d(T=300\,K)\approx 1m$s of the $^{15}$N-V$^{-}$ center. 
\end{abstract}

\maketitle

\begin{figure}[!htb]
\begin{center}
  \includegraphics[width=9cm]{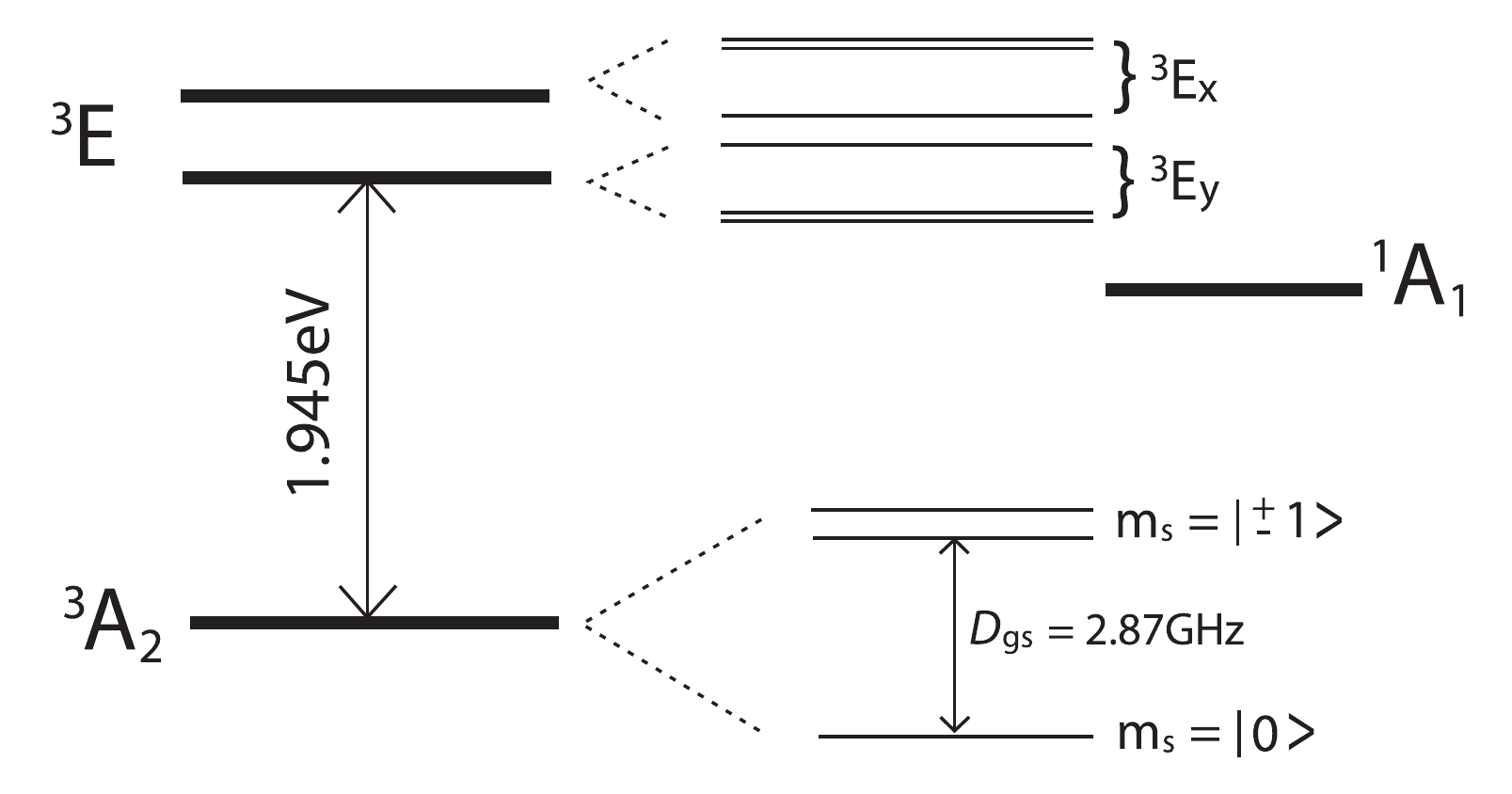}
  \caption[]{Schematic diagram for the energy levels of an NV$^{-}$ center depicting the spin triplets for the ground state $^{3}$A$_2$ and the excited state $^3$E. The splitting due to spin orbit interaction is shown for the excited state.} 
\label{Nvenergy}
\end{center}
\end{figure}

\begin{figure}[!ht] 
\begin{center}
  \includegraphics[height=5cm,width=8.5cm]{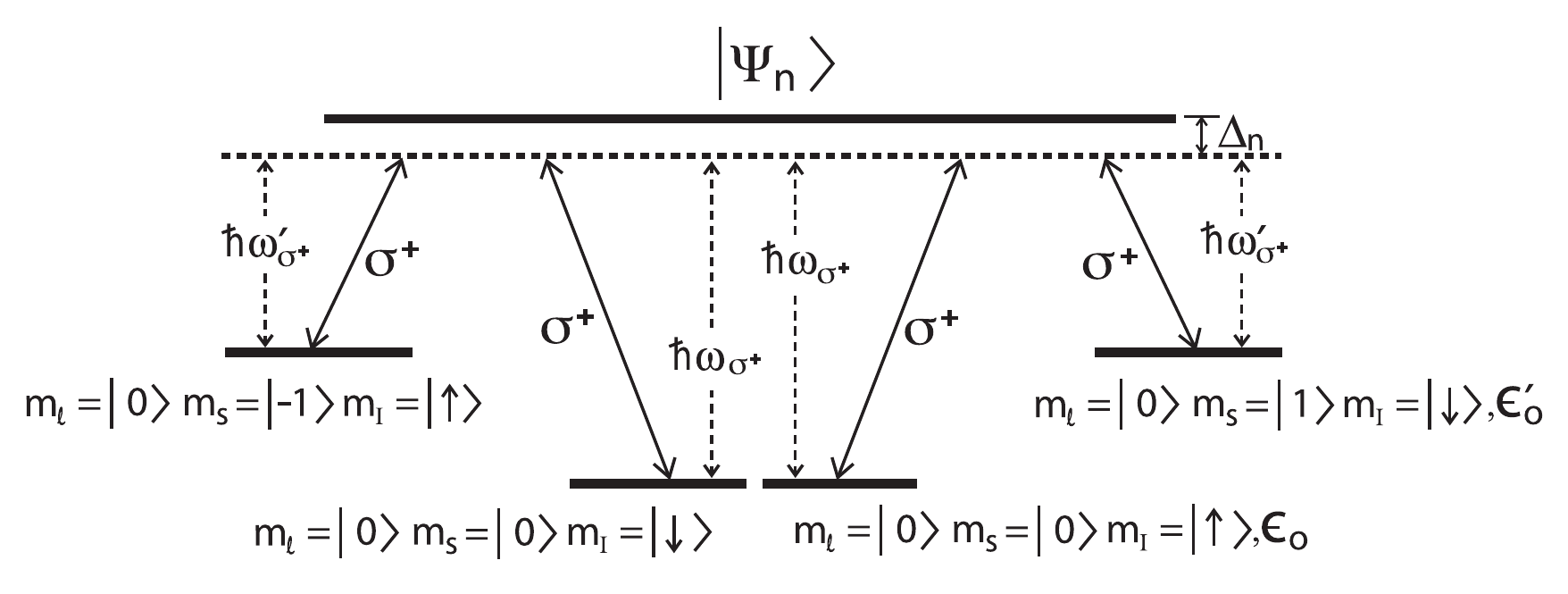}
  \caption{The $\Lambda$ transition optically driven with $\sigma^{+}$-polarized laser beams will flip the nuclear spin projection in order for total angular momentum to be conserved. The $\Lambda$ transition will equally populate both ground state levels due to the two-fold degeneracy between them.} 
\label{rewr}
\end{center}
\end{figure}

Recently tremendous attention has been drawn by single Nitrogen-vacancy defect centers (N-V centers) in diamond due to their promising properties for the experimental realization of quantum computation \cite{1}. In particular, the N-V center in diamond is well suited for studying electronic and nuclear spin phenomena, since its spin can be both initialized and read out via a stable optical spin conserving transitions \cite{1a,1b}. The N-V center appears in two forms, neutral and negatively charged, and it is the negatively charged center (N-V$^{-}$) that we are interested with in this rapid communication. The N-V$^{-}$ stands out among solid-state systems because its electronic spin can be efficiently prepared, manipulated and measured with optical and microwave excitations. Recent experiments have conclusively demonstrated that the N-V center in ultrapure diamond lattice shows the longest room temperature spin dephasing time ever observed in solid state systems (T$_2$=1.8~$m$s) ~\cite{4a}. The electron spin coherence lifetime in the N-V$^{-}$ center is limited by its hyperfine interaction with the Carbon lattice in the diamond structure \cite{4b}. At very low concentration of paramagnetic impurities the loss of electron spin coherence is minimized and spin quantum effects could be observed at room temperature over hundred of nanometers, these results open the door to coherent manipulation of individual isolated nuclear spins in a solid state environment at room temperature with potential applications to quantum computation, therefore the understanding of the hyperfine interaction in this solid state system is of great interest.  \\
In this rapid communication we show how the hyperfine interaction gives rise to a spin non-conserving two-photon $\Lambda$ transition with a rate of the order of 10 MHz at room temperature, which allows for approximately $10^4$ optically controlled quantum logic operations within the spin coherence time at room temperature. The $\Lambda$ transition can be understood by considering the hyperfine interaction of the electron's spin in the N-V$^{-}$ center with the spin-1/2 nucleus of the isotope Nitrogen $^{15}$N. A quantitative result for the number of quantum logic operations is obtained by using second order perturbation theory and numerical diagonalization of the Hamiltonian describing the excited energy levels of the $^{15}$N-V$^{-}$ center when an external electric field is applied. Our results clarify the origin of the two-photon transitions that have been experimentally observed in Refs.~\cite{Santori_OE,Santori_SPIE}. \\
The N-V center in diamond consists of a substitutional nitrogen atom next to a Carbon vacancy giving a centre with C$_{3v}$ symmetry. In the negatively charged state, the extra electron is located at the vacancy site forming a spin $S=1$ pair with one of the vacancy electrons. The energy level structure of the N-V$^{-}$ center was established by combining optical electron paramagnetic resonance and theoretical results. Lenef and Rand earlier work gave a group theoretical description of the molecular orbitals and state configuration of the N-V$^{-}$ center by using linear combinations of $sp^3$ orbitals \cite{7}. Recent numerical {\it ab intio} calculations conducted by Hossain {\it et al}. indicated the following orthonormal orbitals which irreducibly transform according to the C$_{3v}$ symmetry group :
$u = s_{N}$,  $v = (s_{1}+s_{2}+s_{3}-3\beta s_{N})/\sqrt{3}S_{1}$,  
$e_{x}=(2s_{3}-s_{1}-s_{2})/\sqrt{3}S_{2}$, $e_{y}=(s_{1}-s_{2})/S_{2}$, 
where $s_{N}$, $s_{1}$, $s_{2}$ and $s_{3}$ denote $sp^3$ orbitals pointing to the Nitrogen and three Carbons neighboring the vacancy, $S_{1}=\sqrt{1+2\alpha-3\beta^2}$, $S_{2}=\sqrt{2-2\alpha}$, $\alpha=\langle s_1|s_2\rangle$ and $\beta=\langle s_3|u\rangle$ are normalization constants \cite{8}.  
The energy levels are traditionally labeled according to the orbital and spin multiplicity, which is related to the number of states having the same energy. For the negatively
charged N-V center, the ground state is a spin triplet state $^{3}$A$_2$, with a zero-field
splitting of $D_{gs}=2.87$ GHz between spin sublevels $m_s=0$
and $m_s=\pm 1$ [see Fig. \ref{Nvenergy}]. 
\\ 
The excited state $^{3}$E is also a spin
triplet, associated with a broadband photoluminescence
emission with zero phonon line at $1.945$ eV, which allows
optical detection of single N-V defects using confocal
microscopy. It is now well established that a metastable state $^1$A$_1$ is place between the ground and excited triplet states \cite{9}.  
Although an extensive documentation have been made to unveil the excited state structure of the N-V center from first principles, it remains a topic of current research due to its possible quantum applications \cite{10}. \\
The Hamiltonian suitable to describe the $^{3}$E excited state of the negatively charged $^{15}$N-V$^{-}$ center is given by \cite{11}
\begin{equation}
{\cal H}\!\!=\!\!H_{el}+\lambda{\bf L}\cdot{\bf S}+\rho[({\bf L}\cdot{\bf S})^{2}+\frac{1}{2}{\bf L}\cdot{\bf S}-\frac{1}{3}L(L+1)S(S+1)]+A{\bf I}\cdot{\bf S}.
\label{htotal}
\end{equation}
The first term in Eq. (\ref{htotal}) is the dominant term in the Hamiltonian and is given by $H_{el}={\bf F}\cdot{\bf D}$, where ${\bf F}$ is the external applied electric field in the N-V defect and ${\bf D}$ is the dipole moment operator, respectively. From earlier work done by Hossain {\it et al}. we know that the only nonzero dipole moments components are ${\bf D}_{e_{x}}$ and ${\bf D}_{e_{y}}$ with equal dipole magnitudes, i.e. $|{\bf D}_{e_{x}}|=|{\bf D}_{e_{y}}|=D$, respectively.  The dipole moment transverse terms cause a linear splitting of the degenerate ${^{3}}$E state levels given by $\pm D\sqrt{(F_{x}+f_{x})^2+(F_{y}+f_{y})^2}$ \cite{11}, where we have taken into account that the presence of strain has the same effect as a weak electric field denoted by ${\bf f}$. The second and third terms in Eq. (\ref{htotal}) describe the spin-orbit and the spin-spin interaction in the $LS$ coupling approximation \cite{book}, respectively. \\
With this information and restricting ourselves to the $m_{l}=~\pm~\!1$ state sublevels of the spin triplet we can write the electric field, the spin-orbit and spin-spin interactions in the following matrix form  
\begin{widetext}
%\scriptsize
\begin{equation}
H_{el}+H_{ss}+H_{so}=\left(  
\begin{array}{cccccc}
-\lambda+\frac{13}{6}\rho & iD(f_{y}+F_{y}) & 0 & 0 & 0 & \rho \\
-iD(f_{y}+F_{y}) & \lambda+\frac{13}{6}\rho & 0 & 0 & 0 & 0  \\
0 & 0 & -\frac{1}{3}\rho & iD(f_{y}+F_{y}) & 0 & 0  \\
0 & 0 & -iD(f_{y}+F_{y}) & -\frac{1}{3}\rho & 0 & 0  \\
0 & 0 & 0 & 0 & \lambda+\frac{13}{6}\rho & iD(f_{y}+F_{y}) \\
\rho & 0 & 0 & 0 & -iD(f_{y}+F_{y}) & -\lambda+\frac{13}{6}\rho   \\
\end{array} 
\right)\otimes\left( 
\begin{array}{cc}
1 & 0 \\
0 & 1 \\
\end{array} \!\!\right),
\label{HM1}
\end{equation}
\end{widetext}
where we have chosen the basis $|m_{l}m_{s}\rangle$ in the following order $[|-11\rangle,|-1-1\rangle,|-10\rangle,|10\rangle,|11\rangle,|1-1\rangle]$ for numerical purposes and we have assumed that the electric field is applied in the $y$ direction. The direct product between matrices takes into account the nuclear spin subspace, i.e. $m_{I}=\pm 1/2$.  \\
One can see that the spin-orbit interaction does not couple terms with different orbital projections and produces just an energy splitting between the states $m_{s}=\pm 1$. The spin-spin interaction however couples the first and the last state with different orbital projections and produces an energy splitting between the states $m_{l}=\pm 1$. The Hamiltonian given in Eq.~(\ref{HM1}) can be solved exactly and it is easy to convince oneself from the exact solution that there is no overlapping between eigenstates with different spin projections.
Hence, the spin-orbit and spin-spin interaction cannot account for 
the mixing of the excited states and therefore cannot give rise to spin non-conserving transitions. If we want to allow for spin non-conserving transitions between one ground state and another via a common excited state we must introduce the nuclear-spin interaction. The hyperfine interaction will mix the eigenstates with different spin projections allowing us to have a transition to two different ground states. This particular transition is a second order process that can be calculated by using Fermi's golden rule and is commonly known as a $\Lambda$ transition. Therefore, the last term in the total Hamiltonian given in Eq.~(\ref{htotal}) is the nuclear-spin interaction and this term will be responsible for the 
mixing of the excited states, which will give rise to $\Lambda$ transitions in the $^{15}$N-V$^{-}$ center in diamond. The hyperfine interaction can be written in the above basis as
\begin{equation}
H_{ns}\!\!=\!\!A\!\!\left(\!\!   
\begin{array}{cccccccccccc}
\frac{1}{2} & 0 & 0 & 0 & 0 & 0 & 0 & 0 & 0 & 0 & 0 & 0 \\
0 & -\frac{1}{2} & 0 & 0 & \sqrt{2} & 0 & 0 & 0 & 0 & 0 & 0 & 0 \\
0 & 0 & -\frac{1}{2} & 0 & 0 & \sqrt{2} & 0 & 0 & 0 & 0 & 0 & 0 \\
0 & 0 & 0 & \frac{1}{2} & 0 & 0 & 0 & 0 & 0 & 0 & 0  & 0 \\
0 & \sqrt{2} & 0 & 0 & 0 & 0 & 0 & 0 & 0 & 0 & 0 & 0 \\
0 & 0 & \sqrt{2} & 0 & 0 & 0  & 0 & 0 & 0 & 0 & 0 & 0 \\
0 & 0 & 0 & 0 & 0 & 0 & 0  & 0 & 0 & \sqrt{2} & 0 & 0 \\
0 & 0 & 0 & 0 & 0 & 0 & 0 & 0 & 0 & 0 & \sqrt{2}& 0 \\
0 & 0 & 0 & 0 & 0 & 0 & 0 & 0 & \frac{1}{2} & 0 & 0& 0 \\
0 & 0 & 0 & 0 & 0 & 0 & \sqrt{2} & 0 & 0 & -\frac{1}{2}& 0 & 0 \\
0 & 0 & 0 & 0 & 0 & 0 & 0 & \sqrt{2} & 0 & 0& -\frac{1}{2} & 0 \\
0 & 0 & 0 & 0 & 0 & 0 & 0 & 0 & 0 & 0 & 0 & \frac{1}{2}
\end{array} 
\!\!\right),
\label{HM}
\end{equation}
where $A$ is the excited state hyperfine coupling constant which for the $^{15}$N nuclear spin case is $A\approx 60$ MHz \cite{A}. 
\begin{figure}[htb]
\begin{center}
  \includegraphics[width=9cm]{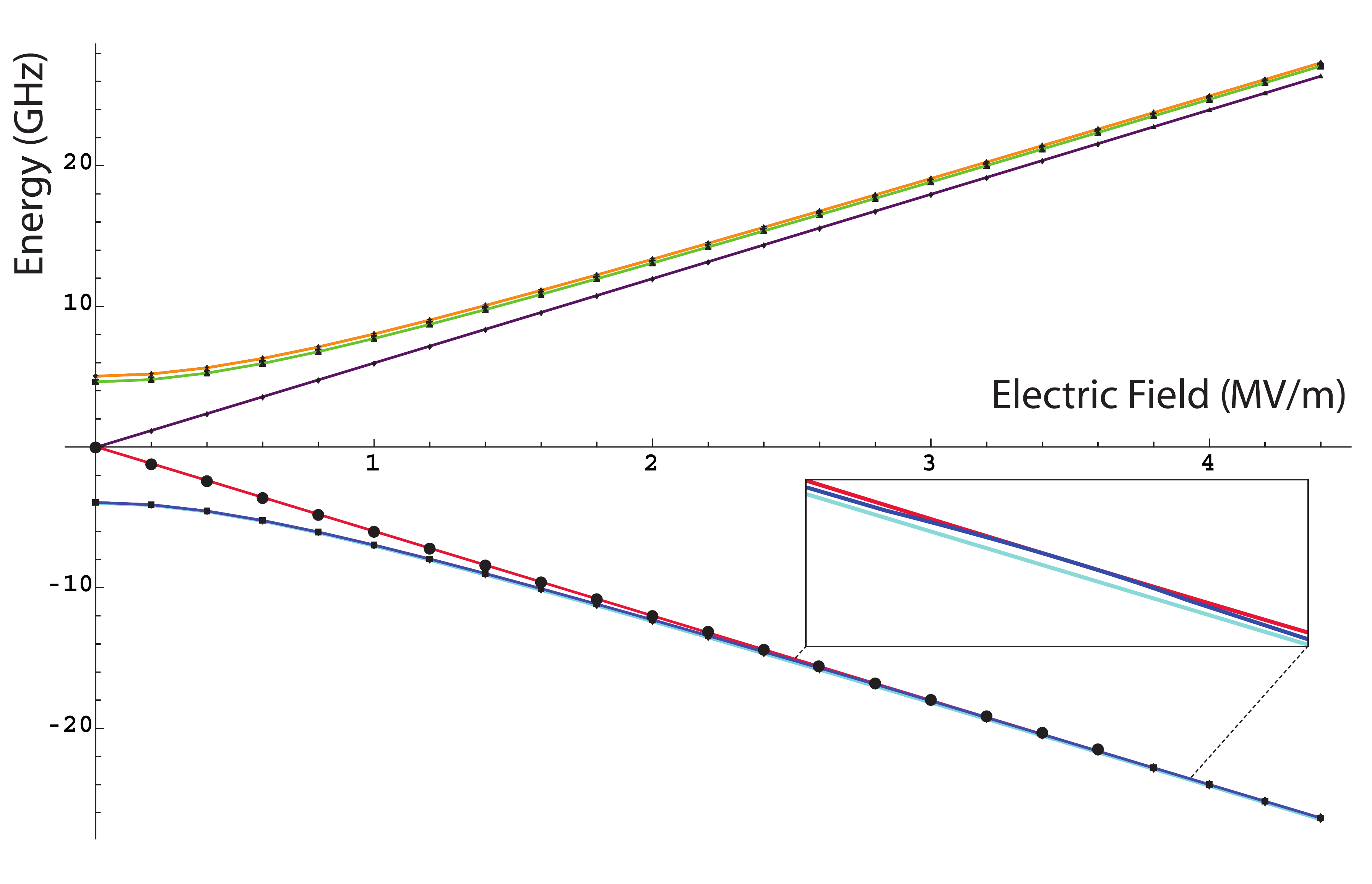}
  \caption[]{Energy diagram showing the eigenvalues of the total Hamiltonian given in Eq. (\ref{htotal}) as a function of the applied external electric field. The lower branch shows the mixing between the states $m_{s}=\pm 1$. The eigenvalues are doubly degenerate.} \label{Energy}
\end{center}
\end{figure}
The eigenvalues of the total Hamiltonian given by Eq. (\ref{htotal}) are plotted as a function of the electric field in Fig.~\ref{Energy}. The energy diagram in Figure~\ref{Energy} shows the mixing between the states $m_{s}=\pm 1$. This mixing will allow us to induce $\Lambda$ transitions between these states in order to write quantum information by using circularly polarized light.
The $^{15}$N-V$^{-}$ center has optical transitions between the ground state $^{3}$A$_2$ and the excited state $^3$E at zero magnetic field that are spin conserving and will allow the readout of information through the electron's spin \cite{11}. In order to write information we need to have optical transitions from two ground state levels via a common excited state. 
The induced transition probabilities for the case of circularly polarized light around the $z$-axis is denoted by $\sigma^{\pm}$ and is proportional to the modulus squared of the matrix element given by $\langle \psi_{i}|\sigma^{\pm}|\psi_{j}\rangle$,
where $\psi_{i}$ and $\psi_{j}$ are electronic wave functions for the initial and final state, respectively, and $\sigma^{\pm}=\mp\sum_{n}(x_{n}\pm iy_{n})/\sqrt{2}$, where the summation are over the electrons in the N-V$^{-}$ center.\\
Suppose that in the absence of light the $^{15}$N-V$^{-}$ system is prepared in the ground state $|m_{l}m_{s}\rangle|m_{I}\rangle=|00\rangle|\uparrow\rangle$. If a strong external electric field with magnitude $F_{0}$ is applied and the crossing between different states occur at a given energy $\epsilon_{n}$ for the n$^{th}$ excited state as illustrated in Fig.~\ref{Energy}, then the ground state will be coupled to an excited state given by
\begin{equation}  
|\Psi_{n}\rangle=\sum_{m_{l}=\pm1,m_{s}=0,\pm1,m_{I}=\uparrow,\downarrow}C^{n}_{m_{l}m_{s}m_{I}}|m_{l}m_{s}m_{I}\rangle
\label{vstates}
\end{equation}
The coherent superposition of states given in Eq. (\ref{vstates}) allows the possibility to have optical transitions between two ground states with different spin projections, i.e. spin non-conserving transitions, by using either a right or left circularly polarized light. In order for the final total angular momentum to equal the initial angular momentum, the nuclar spin projection flips [See Fig.~\ref{rewr}]. The amplitude ${\cal A}_{\pm}$ for the $\Lambda$ transition in the dipole approximation from the initial state $|00\rangle|\uparrow\rangle$ to the final state $|01\rangle|\downarrow\rangle$ through either right or left circularly polarized light is given in second order perturbation theory by
\begin{equation}
{\cal A}_{\pm}=e^{2}{\cal E}^{2}\sum_{n=1}^{12}\frac{\langle\downarrow|\langle 01|\sigma^{\pm}|\Psi_{n}\rangle\langle\Psi_{n}|\sigma^{\pm}|00\rangle|\uparrow\rangle}{\omega_{\sigma^{\pm}}-(\epsilon_{n}-\epsilon_{0})},
\label{amp}
\end{equation}
where $e$ is the electron charge, ${\cal E}$ is the amplitude of the drive optical field, $\epsilon_{0}$ is the energy of the initial ground state and $\omega_{\sigma^{\pm}}$ is the frequency of the circularly right or left polarized light, respectively. 
Writing the explicit form of $|\Psi_{n}\rangle$ in terms of the basis vectors in Eq. (\ref{amp}) we get 
\begin{equation}
{\cal A}_{\pm}=e^{2}{\cal E}^{2} \sum_{n=1}^{12}\frac{C^{n}_{\mp11\downarrow}\sigma^{\pm}_{0,\mp1}C^{n\ast}_{\pm10\uparrow}\sigma^{\pm}_{\pm1,0}}{\omega_{\sigma^{\pm}}-(\epsilon_{n}-\epsilon_{0})},
\label{amp1}
\end{equation}
where $\langle i|\sigma^{\pm}|j\rangle=\sigma^{\pm}_{i,j}$. The value of the $C^{n}_{m_{l}m_{s}m_{I}}$ coefficients and eigenvalues $\epsilon_{n}$ can be obtained by numerical diagonalization of the total Hamiltonian for a given value of the applied external electric field $F_0$, [see Table \ref{coeff}]. 
\begin{table}[!ht] 
\renewcommand{\arraystretch}{1.5}
%\scriptsize
\centering
\begin{center}
\begin{tabular}{|c | c | c | c | c | c|} 
\hline
n & $\epsilon_{n}$(GHz) & $C^{n}_{-11\downarrow}$ & $C^{n}_{-10\uparrow}$ & $C^{n}_{10\uparrow}$ & $C^{n}_{11\downarrow}$ \\ [0.5ex]
\hline 
1 & 12.8540 & 0.5633 & 0.0056 & -$i$0.0021 & $i$0.3965 \\ \hline
2 & 12.8540 & -0.1011 & -0.0010 & -$i$0.0004 & -$i$0.0712 \\ \hline 
3 & 12.4179 & 0.5754 & 0.0292   & -$i$0.0285 & -$i$0.3770 \\ \hline
4 & 12.4179 & -0.1540 & -0.0078  & -$i$0.0076 & $i$0.1009 \\ \hline
5 & -11.3719 & -0.1261 & 0.08755 & $i$0.0879 & -$i$0.2099 \\ \hline
6 & -11.3719 & -0.3276 & 0.2274  & $i$0.2282 & -$i$0.5452 \\ \hline
7 & -11.1861 & -0.2090 & -0.6498 & -$i$0.6502 & -$i$0.1005 \\ \hline
8 & -11.1861 & -0.0050 & -0.0157 & -$i$0.0157 & -$i$0.0024 \\ \hline
9 & -11.1448 & -0.3864 & 0.1344  & $i$0.1306 & $i$0.5767 \\ \hline
10 & -11.1448 & 0.0243 & -0.0084 & -$i$0.0082 & -$i$0.0362 \\ \hline
11 & 10.9909 & 0.0047 & -0.1201  & $i$0.1201 & -$i$0.0021 \\ \hline
12 & 10.9909 & -0.0271 & -0.6961 & -$i$0.6962 & $i$0.0122 \\ 
\hline 
\end{tabular}
\end{center}
\caption{Numerical values for the coefficients $C^{n}_{m_{l}m_{s}m_{I}}$ and eigenvalues $\epsilon_{n}$ of the excited state levels that contribute to the transition amplitude ${\cal A}_{\pm}$ for the case $F_0=1.85$ MV/m \cite{11a}.} \label{coeff}
\end{table}
In the subsequents calculations we will work with right circularly polarized only. In order to calculate the transition rate between the initial ground state and the final ground state we use Fermi's golden rule
\begin{equation}
W_{|00\rangle|\uparrow\rangle\rightarrow|01\rangle|\downarrow\rangle}=\frac{2\pi}{\hbar^4}\left|{\cal A}\right|^{2}\rho(\omega),
\label{Fermi}
\end{equation}
where $\rho(\omega)$ represents the allowed photon frequencies and can be written in its simplest form as a Lorentzian function
\begin{equation}
\rho(\omega)=\frac{(\gamma/2\pi)}{[\omega_{\sigma^{+}}-\omega^{\prime}_{\sigma^{+}}-(\epsilon_{0}^{\prime}-\epsilon_{0})]^2+(\gamma/2)^2},
\label{lo}
\end{equation}
where $\epsilon_{0}^{'}$ is the energy of the final ground state and $\gamma$ is the decoherence rate. For the resonant case, i.e. $\omega_{\sigma^{+}}-\omega^{\prime}_{\sigma^{+}}=(\epsilon_{0}^{'}-\epsilon_{0})$, Fermi's golden rule is given by
$W_{|00\rangle|\uparrow\rangle\rightarrow|01\rangle|\downarrow\rangle}=4|{\cal A}|^2/\gamma\hbar^4$.
Taking the value of the dipole moment matrix elements as $e\sigma^{+}_{0,-1}=e\sigma^{+}_{1,0}=|d|$ in Eq. (\ref{amp1}) we get  
\begin{equation}
{\cal A}=\hbar^2\Omega_{R}^{2} \sum_{n=1}^{12}\frac{C^{n}_{-11\downarrow}C^{n\ast}_{10\uparrow}}{\Delta_{n}},
\label{amp2}
\end{equation}
where $\Delta_{n}=\omega_{\sigma^{+}}-(\epsilon_{n}-\epsilon_{0})$ is the detuning and we have used the relation ${\cal E}|d|=\hbar\Omega_{R}$ where $\Omega_{R}$ is the Rabi frequency. 
In order for perturbation theory to be valid we require that $|\Delta_{n}|\gg\Omega_{R}|C^{n\ast}_{10\uparrow}|$, hence we have to tune the frequency $\omega_{\sigma^{+}}$ to achieve this. Assuming that the detuning is taken in such a way that $|\Delta_{n}|\approx30\Omega_{R}|C^{n\ast}_{10\uparrow}|$, where the Rabi frequency is a few GHz \cite{pentagon}, then the amplitude is $|{\cal A}|^2\approx 10^{-27}\hbar^{4}\Omega^{4}_{R}$. Using the spin coherence time of $\gamma^{-1}\approx1 m$s \cite{4a}, we can estimate the $\Lambda$ transition rate to be of the order of
\begin{equation}
W_{|00\rangle|\uparrow\rangle\rightarrow|01\rangle|\downarrow\rangle}\approx 10\,\,\mbox{MHz}.
\label{result}
\end{equation}
Based on the obtained transition rate and for coherence lifetimes of around the milliseconds \cite{4a}, we estimate that the $^{15}$N-V$^{-}$ center in diamond should be capable of around $10^{4}$ quantum operations per coherence lifetime at room temperature, which confirms that this solid state system is a very good candidate for quantum computing \cite{hemm}.\\
Suppose now that we can prepare the $^{15}$N-V$^{-}$ center in such a way that all the spin population is in the ground state level $m_{s}=|0\rangle$. We have shown that an optically driven $\Lambda$ transition will change the nuclear spin projection because of the spin non-conserving transition. Therefore if we optically pump the ground state level so that we have a $\Lambda$ transition of the form $|m_s=0\rangle\rightarrow|m_s=\pm1\rangle$, then the ground state population will transfer according to the two possible $\Lambda$ transitions $|m_s=0\rangle|\uparrow>\rightarrow|m_s=\pm1\rangle|\downarrow>$ or $|m_s=0\rangle|\downarrow>\rightarrow|m_s=\pm1\rangle|\uparrow>$, with flipped nuclear spin projection [see Fig.~\ref{rewr}]. Note that both $\Lambda$ transitions are completely independent from each other. 
Then, by means of resonance condition, it is possible to produce an arbitrary superposition of the form $|\Psi_{w}\rangle=\alpha|m_{s}=0\rangle+\beta|m_{s}=1\rangle$, which represents the qubit. 
  
In order to read this superposition state the electric field $F_y$ needs to be switched off, leading to ${\cal A}=0$. This means that for $F_y=0$ and weak strain the two-photon transitions are spin conserving, which allows for the detection e.g. $|\alpha|^2$ through the linear response to a $\sigma^+$-polarized laser beam. Single-qubit operations of the qubit state $|\Psi_{w}\rangle$ can be performed by means of the spin conserving and spin non-conserving transitions, which give rise to phase and amplitude shifts, respectively.\\
In conclusion, we have demonstrated that there exists a $\Lambda$ transition in $^{15}$N-V$^{-}$ center in diamond that is mediated by the hyperfine interaction which allows us to write quantum information in this solid state system. Readout can be performed when the electric field is tuned to spin conserving transitions. Using second order perturbation theory we have obtained a $\Lambda$ transition rate of the order of 10 MHz at room temperature, which allows for approximately $10^4$ quantum logic operations within the spin coherence time $\tau_d(T=300\,K)\approx 1m$s of the $^{15}$N-V$^{-}$ center. Our findings clarify the origin of the two-photon transitions observed in Refs.~\cite{Santori_OE,Santori_SPIE}, which is essential for the development of a scalable quantum network made of $^{15}$N-V$^{-}$ centers in diamond, given the possibility to produce a photonic crystal structure in diamond \cite{Awschalom,Barth}.\\ \\
%%%%%%%%%%%%%%%%%%%%%%%%%%%%%%%%%%%%%%%%%%%%%%%%%%%%%%%%%%%%%%%%%%%%%%%%%%%%%
We acknowledge support from NSF ECCS-0725514, DARPA/MTO HR0011-08-1-0059, NSF ECCS-0901784, and AFOSR FA9550-09-1-0450.
%%%%%%%%%%%%%%%%%%%%%%%%%%%%%%%%%%%%%%%%%%%%%%%%%%%%%%%%%%%%%%%%%%%%%%%%%%%%

%%%%%%%%%%%%%%%%%%%%%%%%%%%%%%%%%%%%%%%%%%%%%%%%%%%%%%%%%%%%%%%%%%%%%%%%%%%%

\end{document}